# Robust Audio Watermarking Using Graph-based Transform and Singular Value Decomposition

Majid Farzaneh, Rahil Mahdian Toroghi
Media Engineering Faculty, Iran Broadcasting University, Tehran, Iran
Majid.farzaneh91@gmail.com , Mahdian.t.r@gmail.com

*Abstract*— **Graph-based Transform (GT) has been recently leveraged successfully in the signal processing domain, specifically for compression purposes. In this paper, we employ the GBT, as well as the Singular Value Decomposition (SVD) with the goal to improve the robustness of audio watermarking against different attacks on the audio signals, such as noise and compression. Experimental results on the NOIZEUS speech database and MIR-1k music database clearly certify that the proposed GBT-SVD-based method is robust against the attacks. Moreover, the results exhibit a good quality after the embedding based on PSNR, PESQ, and STOI measures. Also, the payload for the proposed method is 800 and 1600 for speech and music signals, respectively which are higher than some robust watermarking methods such as DWT-SVD and DWT-DCT.**

## I. Introduction

Ever since the transmission, production, and releasing of multimedia contents (e.g., video, audio, and image) through internet has been simplified, the basic need for data protection against copying and unauthorized distribution has been increased, accordingly. These concerns have prompted researchers to look for reliable ways to deal with the copyright protection. In this regard, the most promising solutions introduced so far are based on hiding information algorithms [1].

Hiding information is a process, in which a message is embedded in a digital media. The embedded message should be invisible (in a video, or an image) or inaudible (in a sound). In addition, the main media should remain as original. In other words, the embedding of the message should not make tangible changes in the main media [1]. Hiding information could be divided into two categories: Steganography and Watermarking. The main purpose of steganography is to hide the principle of communication. The transmitter embeds a serial message in a digital media (such as audio), where only the receiver is able to extract it.

Watermarking is pretty similar to steganography, since both aim at hiding the information. However, the steganography is used for point-to-point communication between two parties. Therefore, Steganography usually has a limited resistance to the changes and effects which might occur for the transmitted signal due to formatting, compression, or even analog-to-digital conversion.

In contrast, Watermarking uses different rules. When media (i.e., audio or image) is available to the people who are aware of the presence of some hidden information in its contents, they may elaborate to get access to that information. Hence, the issue of resisting against attacks is very important in Watermarking. The most important applications of watermarking could be summarized in the following three categories [2]:

1) Copyright Protection (copyright): This aims at protecting the rights of the authors. In this application, the owner's information is embedded in the media [3].

2) Markup: In this application, something similar to the serial number is embedded in a master copy of a digital file in order to identify the original copies of the counterfeit [4].

3) Content Authentication: In this application, we can see if a digital file has been manipulated [5].

The techniques and applications of watermarking described in previous studies have been more focused on images, the audio contents have been less considered, whereas many readers or audiovisual owners such as audio books or online radios, suffer from copyright infringement.

Recently, graph-based signal processing techniques have gained the attention of researchers. One of the applications of graphical processing is the graph-oriented conversion, which is often used to compress information [6, 7].

A method for audio compression by Graph-based Transform is reported in [8], which proposes this method over a popular conventional method, namely DCT (discrete cosine transformation). The main idea of this work is to use this method to add the watermark data to parts of the audio signal which are resistant to compression. If the watermark data is added to the important parts of the signal, it will almost remain unchanged during compression, changing the format, or corruption by additive noise. In this paper, a graph-based transformation technique is developed for audio signals, which improves the robustness of watermarking against various attacks.

The rest of the paper is organized as follows:
Section II describes the Graph-based Transform, Section III introduces the proposed GBT-SVD audio watermarking method, Section IV reports the experimental results, and Section V provides the conclusion.

## II. Graph-based Transform (GBT)

Given a block of an audio signal with a frame size of *N* samples, we can create a graph *G={V,E,s}* where *V* and *E* are the vertices and edges of the graph, and $s \in \mathbb{R}^{N \times 1}$ is an audio signal for which the graph matrix is defined as $K \in \mathbb{R}^{N \times N}$. For this graph, the adjacency matrix A, elements are obtained as

$$A_{ij} = \begin{cases} a_{ij}, & if\ (i.j) \in E \\ 0, & otherwise \end{cases} \quad (1)$$

Where $a_{ij}$ is the weight of the edge between *i* and *j* in the graph. The degree matrix $D \in \mathbb{R}^{N \times N}$ is a diagonal matrix, for which the elements are defined as follows,

$$D_{ij} = \begin{cases} \sum a_{ij}, & if\ i = j \\ 0, & otherwise \end{cases} \quad (2)$$

Then, the Graph-Laplacian Matrix *L* would be defined as,

$$L = D - A \quad (3)$$

Where the operator *L* is also known as *Kirchhoff* operator, as a tribute to Gustav Kirchhoff for his achievements on electrical networks. Kirchhoff referred to the (weighted) adjacency matrix *A* as the *conductance* matrix.

The matrix L would be a real symmetric one and based upon the spectral theory, the eigenvalue decomposition (EVD) of this matrix would lead to a set of real non-negative eigenvalues, denoted by $\Lambda = \{\lambda_1, \ldots, \lambda_N\}$, and a set of corresponding independent (hence, orthogonal) eigenvectors denoted by $V = \{v_1, \ldots, v_N\}$, derived as,

$$L = V \Lambda V^T \quad (4)$$

Then we can use these orthogonal eigenvectors to de-correlate the signal defined on the graph, i.e.,

$$c = V^T s \quad (5)$$

Where $c \in \mathbb{R}^{N \times 1}$ is the approximate sparse transform coefficient matrix, [1].

## III. Proposed Audio Watermarking Method

The proposed audio watermarking method is based on GBT and Singular Value Decomposition (SVD). At first, we need to divide the cover audio signal to several frames with equal samples. Each watermark bit should be embedded in one frame. The process is, as follows:

1) To do the embedding, first, we apply the GBT on the audio frame. The GBT coefficients are equal to the number of frame samples. However, instead of using the entire coefficients, we only pick 40% of the first coefficients, since those coefficients consist of the maximum and sufficient information for the audio frame to perform the embedding. This will help us have a bigger quality for the audio signal after the embedding process. The SVD is applied to those 40% selected coefficients. The decomposition provides us with the largest singular value ($S_{max}$). Then, we embed the watermark bit into the $S_{max}$ as follows:

**if** *Watermark_bit == 0*
  $S_{max} = S_{max} - WS$
**else if** *Watermark_bit == 1*
  $S_{max} = S_{max} + WS$
**end**

Where the value of *WS* is a constant value indicating the Watermarking Strength.

Since the number of frames is usually more than the watermark bits, we prefer to be fastidious about choosing the frames which are more useful for embedding. The frames with larger energy values have larger singular values and are more robust against noise, and therefore are preferred.

The Extraction step is performed to retrieve the watermark image from the Watermarked Audio. To extract the embedded watermark, we repeat the same processing blocks of framing, GBT, and SVD for the watermarked audio as well, and then we compare the largest singular value in the watermarked frames ($S^W_{max}$) with the saved largest singular values in the embedding step ($S_{max}$). Therefore, we can obtain the watermark bits, as

**if** $S^W_{max} > S_{max}$
  *Watermark_bit = 1*
**else**
  *Watermark_bit = 0*
**end**

Figure 1, shows the entire flowchart of the proposed watermarking process including embedding and extracting in the presence of attacks. *FLOWCHART 1* and *FLOWCHART 2* (Figure 2, and Figure 3), show the embedding and extracting steps in details for one bit in one frame, respectively.

For the GBT step, we need to consider an appropriate graph structure. This graph should de-correlate the audio frames and provide some large coefficients at the first samples, while the rest of the coefficients (which are the majority of coefficients) should be very small or even zero.

We showed in [8] that, Figure 4 is an appropriate graph structure for the audio signals and provides a better compression rate than the one used through DCT. Following our work [8], we use the same structure in the present work. Figure 2, shows the appropriate graph structure for a frame with a length of 8 samples. In this structure, every sample is related to the neighboring samples, since we know that in audio signals nearby samples are highly correlated. This structure can lead to a high de-correlation after GBT. The matrix *A* in Eq. (6) is a good example of a corresponding adjacency matrix for the proposed graph structure.

$$A = \begin{bmatrix} 0 & 1 & 0.3 & 0 & 0 & 0 & 0 & 0 \\ 1 & 0 & 1 & 0.3 & 0 & 0 & 0 & 0 \\ 0.3 & 1 & 0 & 1 & 0.3 & 0 & 0 & 0 \\ 0 & 0.3 & 1 & 0 & 1 & 0.3 & 0 & 0 \\ 0 & 0 & 0.3 & 1 & 0 & 1 & 0.3 & 0 \\ 0 & 0 & 0 & 0.3 & 1 & 0 & 1 & 0.3 \\ 0 & 0 & 0 & 0 & 0.3 & 1 & 0 & 1 \\ 0 & 0 & 0 & 0 & 0 & 0.3 & 1 & 0 \end{bmatrix} \quad (6)$$

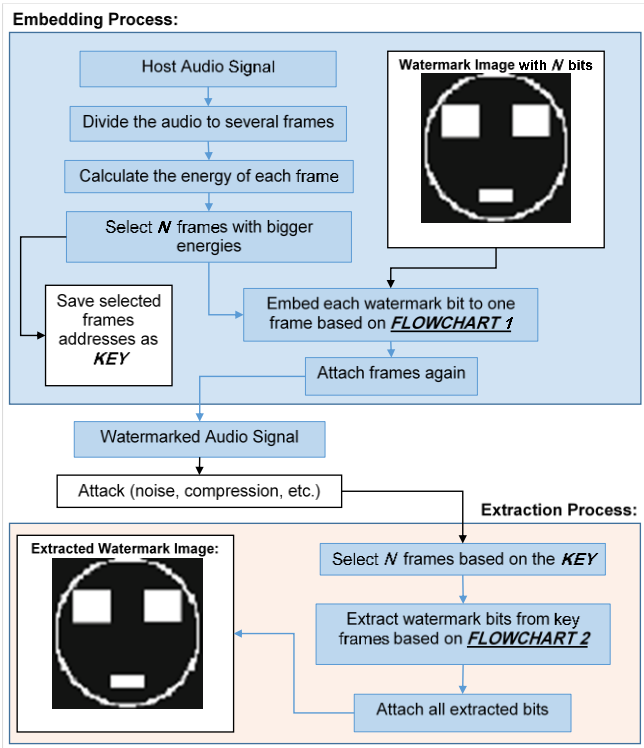

**Fig. 1.** Proposed audio watermarking method consist of embedding and extracting steps

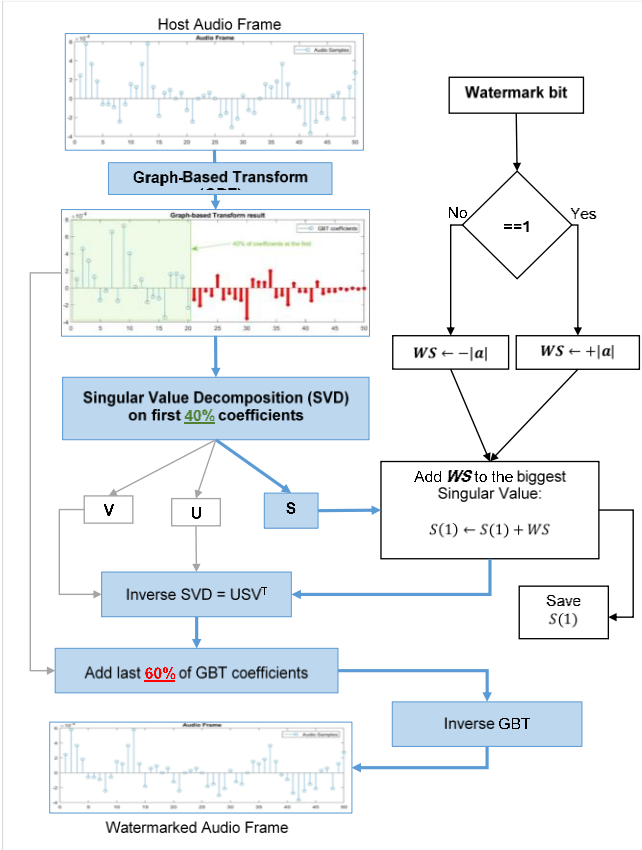

**Fig. 2.** *FLOWCHART 1:* The proposed GBT-SVD-based method to embed one watermark bit into one audio frame

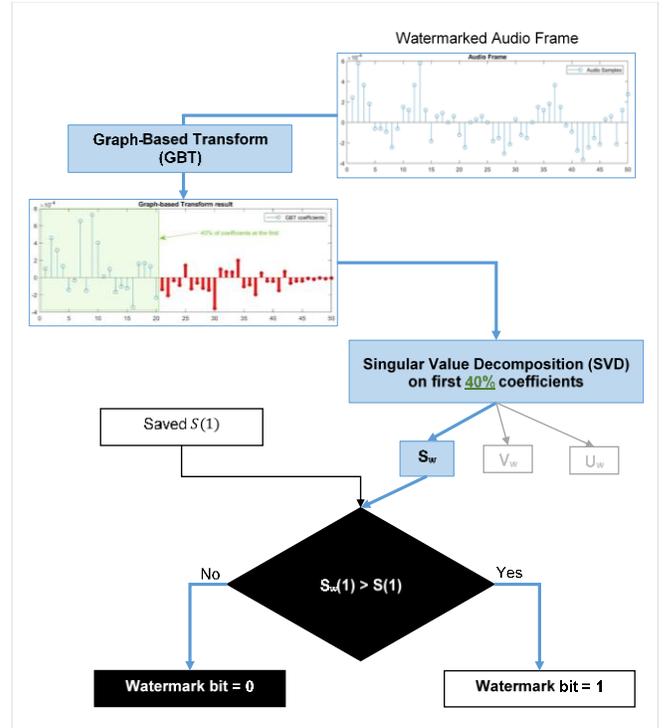

**Fig. 3.** *FLOWCHART 2*: The proposed GBT-SVD-method to extract one watermark bit from a watermarked frame

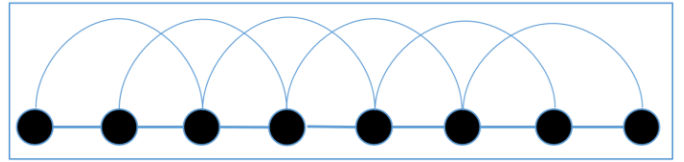

**Fig. 4.** An appropriate graph structure for audio signals according to [8]

## IV. Experimental Results

We have used the NOIZEUS speech database [16] and the MIR-1k music database to test the proposed watermarking method. NOIZEUS consists of 15 male voice signals, as well as 15 female ones. MIR-1k consists of 1000 pop music pieces. To evaluate the quality of audio host signals after the embedding process, we use the Peak Signal to Noise Ratio (PSNR), the Short-Time Objective Intelligibility (STOI) [9], the Perceptual Evaluation of Speech Quality (PESQ) [10] and a Mean Opinion Score (MOS) for 10 audiences between host audio signal and watermarked audio signal. Moreover, to evaluate the extracted watermark images, we use the Bit Error Rate (BER) between the original watermark image and the extracted watermark image. We need to keep PSNR, PESQ, STOI, and MOS high and BER low during the presence of different attacks. The 10 audiences were asked to listen to original audio signals and watermarked audio signals, respectively and give a score between 1 (very annoying) to 5 (no sense of change or embedding a watermark).

To evaluate the robustness of the proposed method, we simulated 8 different attacks including Additive White Gaussian Noise (AWGN), MP3 compression, Low Pass Filtering (LPF), High Pass Filtering (HPF), re-sampling, re-quantization, Amplitude Scaling (AS), and cropping. In our experiments, the SNR values for the AWGN attacks are 20dB and 10dB. The bitrates for MP3 compression are 64 kbps and 32 kbps. Cut-off frequencies for LPF and HPF are 4 kHz and 50 Hz, respectively. In the resampling attack, we down-sample and again up-sample the watermarked signal. The levels for quantization are 24 and 8, and the scaling coefficient for the AS attack is 0.7. For cropping attack, 20% and 30% of the audio signal is removed from the beginning of the signal. For all experiments, we have used two binary images with the size of 25×25 bits. Figure 5, shows these images.

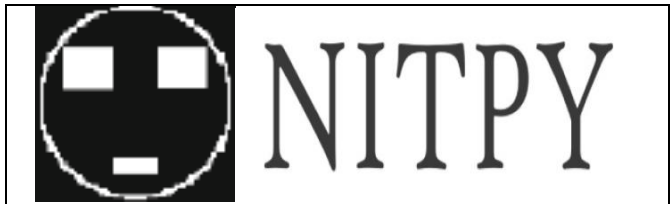

**Fig. 5. Watermark images (one with majority of black pixels and another with majority of white pixels)**

Table1, reports the average values of PSNR, PESQ, STOI, and MOS between the host audio signal and the watermarked audio signal for both speech and music databases. These results clearly show that after embedding a watermark image into the audio signal, the change is not recognizable. The MOS value shows that the audiences could not detect any changes after the embedding process. The watermark strength (WS) value for all test experiments were 0.05, and the frame size was fixed by 10 i.e. each frame has 10 samples.

**Table 1. Average quality measures between the host and the watermarked audio signals**

| Database | PSNR | PESQ | STOI | MOS |
|---|---|---|---|---|
| Speech | 43.26 | 3.46 | 0.93 | 4.9 |
| Music | 50.96 | - | - | 5 |

Table 2, reports the average Bit Error Rates between the original watermark images and the extracted watermark images in the presence of different attacks. We see from the table 2, that the GBT-SVD-based watermarking is robust against most of the attacks. However, it does not seem to be sufficiently robust against re-scaling, whereas it could be more robust by selecting larger values of WS to the cost of losing the quality.

Table 3, provides some comparison with other transform-based audio watermarking methods in the presence of various attacks.

**Table 2. Average BER between original watermark and extracted watermark in presence of various attacks**

| Attack | Proposed on Speech | Proposed on Music | Proposed (average) |
|---|---|---|---|
| No attack | 0.000 | 0.000 | 0.000 |
| AWGN (20 dB) | 0.000 | 0.000 | 0.000 |
| AWGN (10 dB) | 0.000 | 0.097 | 0.048 |
| MP3 (32 kbps) | 0.117 | 0.348 | 0.232 |
| MP3 (16 kbps) | 0.127 | 0.350 | 0.238 |
| Re-sampling (6000) | 0.000 | 0.000 | 0.000 |
| Re-sampling (4000) | 0.001 | 0.000 | 0.000 |
| LPF (4 kHz) | 0.000 | 0.000 | 0.000 |
| HPF (50 Hz) | 0.000 | 0.120 | 0.060 |
| AS (0.9) | 0.024 | 0.556 | 0.287 |
| AS (0.7) | 0.540 | 0.557 | 0.548 |
| Re-quantization (24) | 0.000 | 0.000 | 0.000 |
| Re-quantization (8) | 0.086 | 0.135 | 0.110 |
| Cropping (20%) | 0.143 | 0.109 | 0.126 |
| Cropping (30%) | 0.242 | 0.164 | 0.203 |
| **Average:** | **0.085** | **0.162** | **0.123** |

**Table 3. Comparison BER with other related methods**

| Attack | Proposed | DCT-SVD [12] | DWT-SVD [13] | DWT-LU [14] | DWT-DCT [15] |
|---|---|---|---|---|---|
| AWGN-20 | **0.000** | 0.000 | 0.000 | - | 0.030 |
| MP3-32k | 0.232 | 0.001 | 0.290 | 0.020 | 0.010 |
| Resampling | **0.000** | 0.030 | 0.000 | 0.000 | 0.090 |
| LPF (4 kHz) | **0.000** | 0.000 | 0.189 | 0.030 | 0.040 |
| HPF (50 Hz) | **0.060** | 0.406 | 0.358 | - | - |
| AS (0.7) | 0.548 | 0.31 | - | - | 0.000 |
| Requantization | **0.000** | 0.000 | 0.000 | 0.000 | 0.000 |

The payload in this table, shows the maximum number of bits that can be embedded in one-second of audio. In the proposed method, every frame has 10 samples, and the sampling rate is 8000 Hz for the entire speech database. Thus, for every second we have 800 frames and each frame can hold one bit. Moreover, in the music database, the sampling rate is 16000 Hz. So, the payload for the speech database is 800, while for the music database it is 1600.

Table 4, shows the payload for the proposed method in comparison with the other methods.

**Table 4. Comparing payload with other related methods**

| Method | Payload | Database |
|---|---|---|
| Proposed | 1600 | Music |
| Proposed | 800 | Speech |
| DCT-SVD [12] | 6000 | Both |
| DWT-SVD [13] | 1030 | Both |
| DWT-LU [14] | 1280 | Both |
| DWT-DCT [15] | 86.13 | Music |

As we can see except the AS, the proposed method is robust against all kinds of attacks, and in comparison to other methods, is more robust against AWG noise, resampling, low pass filter, and high pass filter attacks. Against the re-quantization attack, the proposed method is as robust as other methods and against the MP3 compression attack, it is more robust than DWT-SVD [13] and less robust than other methods. By keeping the robustness, the proposed method has the desired quality based on PSNR, PESQ, STOI, and MOS values. Also, it has a good payload that is higher than the payload of DWT-SVD [13], DWT-LU [14] and DWT-DCT [15] methods.

## V. Conclusion

In this paper, we proposed a new robust audio watermarking method based on Graph-based Transform (GBT) and Singular Value Decomposition (SVD). Experimental results show that the proposed method has a high resistance in the presence of various attacks. By selecting 10 samples for each frame and taking watermark strength equal to 0.05 we have a good quality after the embedding process and watermark effect cannot be heard or recognizable. The payload for the proposed method is $f_s/10$ where $f_s$ is the sampling rate frequency of the host audio signal. By considering the accuracy and quality of the method, the payload value is good enough in comparison to other methods.

In future researches, we plan to work on optimal graph structures and analyzing different embedding strategies to achieve more robustness, as well as payload, and quality.